\documentclass[conference]{IEEEtran}
\IEEEoverridecommandlockouts

\usepackage{graphicx}
\usepackage[utf8]{inputenc}  
\usepackage{amsmath}
\usepackage{amssymb}
\usepackage{mathtools}
\usepackage{amsfonts}
\usepackage{gensymb}
\usepackage{microtype}  
\usepackage{pdfpages}
\usepackage[numbers,square]{natbib}

\usepackage{color}
\usepackage{url}
\usepackage{makecell}

\usepackage{authblk}
\usepackage[hidelinks]{hyperref}

\newsavebox\commabox
\sbox{\commabox}{,}

\def\av{{\mathbf a}}

\def\dv{{\mathbf d}}

\def\pv{{\mathbf p}}

\def\xv{{\mathbf x}}

\def\Cm{{\mathbf C}}

\newcommand{\figref}[1]{Figure~\ref{#1}}
\newcommand{\hyp}[1]{$\mathcal{H}_#1$}

\newcommand{\ie}{\emph{i.e.}}
\newcommand{\eg}{\emph{e.g.}}

\title{\textbf{Perceptual Evaluation of Higher-Order Ambisonic Codecs on Both Synthetic Mixing and Native Recordings}}

\author[1]{Adrien Llave\thanks{Submitted to the AES AVARIG Conference 2026 on March 31\textsuperscript{st}, 2026. Correspondence: \href{mailto:adrien.llave@orange.com}{adrien.llave@orange.com}}}
\author[1]{Grégory Pallone}
\author[2]{Jérôme Daniel}

\affil[1]{\textit{\small Orange Research, 35510 Cesson-Sévigné, France}}
\affil[2]{\textit{\small Orange Research, 22300 Lannion, France}}

\date{}

\begin{document}

\maketitle 

\begin{abstract}

Spatial audio is spreading in applications such as virtual and augmented reality and immersive games.
The higher-order ambisonic (HOA) format is particularly useful in this context.
Transmitting spatial information requires multiple channels, \eg, 16 channels for 3\textsuperscript{rd}-order ambisonics, resulting in increased memory requirements for storage and higher bitrates for communication.
Therefore, efficient compression algorithms are necessary for those contents.
The recently standardized IVAS codec allows the coding of HOA content for communication use-cases.
Here, we propose to evaluate it in comparison with a basic multi-mono approach across a variety of contents and spatialization methods.
Results show that IVAS outperforms the multi-mono approach at the same bitrate.
In particular, this codec exploits inter-channel correlation to reduce the bitrate.
We point out that it is therefore especially robust for signals with a high interchannel correlation, such as those composed of a limited number of plane waves.
Conversely, the multi-mono approach is unable to exploit this correlation and performs poorly on this type of signal.
\end{abstract}

\section{Introduction}
For applications such as Virtual and Augmented Reality (AR/VR) and immersive games, spatial audio rendered through headphones has gained increasing importance.
Realism and immersion can be enhanced thanks to head-tracking~\cite{begault_direct_2001}, \ie, by applying a rotation to the auditory scene to compensate for the listener's head movements.
However, performing such rotations based on binaural filters introduces comb-filter artifacts related to interpolation between two filter sets~\cite{lindau_minimum_2008}.
The higher-order ambisonics (HOA)~\cite{daniel_representation_2001} has been adopted for this purpose as it offers a simple solution to this problem.
Indeed, it allows for auditory scene rotation via a simple multiplication of a rotation matrix with the ambisonic signals, known as HOA components~\cite{noisternig_3d_2003, zaunschirm_binaural_2020}.
Furthermore, binaural rendering of the HOA format is typically performed using a fixed filter matrix, which reproduces the wavefront without the spectral coloration artifacts mentioned above~\cite{schorkhuber_binaural_2018}.
Additionally, ambisonic audio recording aligns well with 360\textsuperscript{$\circ$} video capture using compact spherical microphone arrays (SMA).
It has been demonstrated that 3\textsuperscript{rd}-order ambisonics is necessary to achieve accurate localization~\cite{zaunschirm_binaural_2020, huisman_ambisonics_2021}.
However, it entails the transmission of 16 channels in parallel, corresponding to a bitrate of 12,288~kbps for a non-compressed 16-bit quantized signal sampled at 48~kHz.
Therefore, the use of an efficient codec is essential to reduce bandwidth or storage requirements without compromising audio quality.
In particular, for the AR/VR use-cases, the codec must operate with low latency, often referred to as \emph{conversational}.

Since HOA components are audio signals, a straightforward coding strategy consists in applying a mono (or stereo) codec independently to each channel (or each pair of channels, respectively).
This approach is employed by the Opus Channel Family Mapping (CMF)~2 codec~\cite{skoglund_ambisonics_2018, narbutt_streaming_2017}.
Alternatively, Opus CMF~3 proposes to project HOA components onto a grid of virtual loudspeakers, then applying instances of a mono/stereo codec within this representation domain, before reverting to the HOA domain at the decoder~\cite{skoglund_ambisonics_2018, rudzki_improvements_2023}.

It is worth mentioning also MPEG-H~\cite{herre_mpeg-h_2015} which relies on inter-channel correlation analysis.
However, this codec, designed for broadcast/multimedia applications, introduces significant latency, which does not always meet the constraints of real-time applications.

The IVAS codec, the latest standardized conversational codec by 3GPP, includes a Scene-Based Audio (SBA) mode dedicated to HOA format up to 3\textsuperscript{rd}-order~\cite{noauthor_3gpp_2023}.
With a total delay of approximately 38~ms~\cite{weckbecker_ambisonics_2025}, it is a promising candidate for this type of application.
The SBA mode of IVAS combines two methods~\cite{weckbecker_ambisonics_2025}: SPAR~\cite{mcgrath_immersive_2019} and DirAC~\cite{laitinen_reproducing_2011}.
These methods exploit inter-channel correlation to enable efficient compression of the information.
They consist of transmitting a limited number of audio streams, called transport channels (TCs), which are expected to be as decorrelated as possible, along with spatial parameters at a lower temporal resolution.
At decoding, these parameters are used to recombine the TCs to reconstruct the HOA components.
For the 3\textsuperscript{rd}-order ambisonics, IVAS achieves good performance on general audio content but does not reach transparency, even at its maximum bitrate of 512~kbps~\cite{noauthor_3gpp_2024}.
However, its performance has not been characterized in detail according to the type of content and for bitrates below 192~kbps.

Yet, HOA components can be obtained through various methods:
(i) the synthetic spatial encoding of mono sounds weighted by spherical harmonic values corresponding to the desired direction (also called 'plane wave encoding'),
(ii) or thanks to native recording using a microphone array, typically an SMA, \eg, EigenMike 32 and 64 (EM32 and EM64), Zylia (ZM-1), SpaceMic,
(iii) or even by convolving mono sounds with Spatial Room Impulse Responses (SRIRs) corresponding to the desired directions, which can be either measured from SMA recordings or simulated.
In the latter two cases, a spatial encoding step is necessary to derive the HOA components.
Well, the method used to compose the auditory scene can influence the correlation between the HOA components~\cite{hellerud_spatial_2009}.
Since SPAR primarily relies on exploiting the covariance matrix of the signals, different performance levels can be expected depending on the spatialization method and scene composition.

In this work, we compare the performance of the IVAS codec and the naive multi-mono approach by applying EVS codec independently to each channel (EVSx16).
Furthermore, we consider a wide variety of 3\textsuperscript{rd}-order ambisonic content, considering both the types of sound sources and the spatialization methods used.
Specifically, we aim to test the following hypotheses:
(\hyp{1}) IVAS exhibits a bias in favor of spatialized content without diffuse reverb,
(\hyp{2}) this bias is negatively correlated with the number of TCs (and thus the chosen bitrate), \ie, the advantage for synthetic content increases as the number of TCs decreases,
(\hyp{3}) EVSx16 favors spatially diffuse content (low inter-channel correlation),
(\hyp{4}) regardless of bitrate.
The remainder of this manuscript is organized as follows:
Section~\ref{sec:ivas} describes the main principles underlying the compression methods used in IVAS in SBA mode, to allow interpretation of the experimental results.
Section~\ref{ssec:exp1_methods} details the first experiment, which assesses the perceived quality of both codecs at different bitrates across a broad range of content.
Section~\ref{sec:exp2} presents a second experiment that measures more precisely the differences in perceived quality between an auditory scene composed of a limited number of plane waves and another with spatially diffuse reverberation, all other factors being equal.
Finally, Section~\ref{sec:discussion} discusses the results and interprets the observed performances in relation to the coding strategies employed by EVSx16 and IVAS in SBA mode.

\section{IVAS description}\label{sec:ivas}

\begin{table}[b]
    \small
    \centering
    \caption{\label{tab:ivas_bitrate_summary}Summary of the target bitrate allocation between the transport channels (TCs) and the spatial parameters (SP) for IVAS Scene-Based Audio mode.}
    \begin{tabular}{r | r r r r r}
        \hline\hline
        \makecell{Total\\(kbps)} & \makecell{1\textsuperscript{st} TC\\(kbps)} & \makecell{2\textsuperscript{nd} TC\\(kbps)} & \makecell{3\textsuperscript{rd} TC\\(kbps)} & \makecell{4\textsuperscript{th} TC\\(kbps)} & \makecell{SP\\(kbps)} \\
        \hline\hline
         32 & 24 & - & - & - & 8 \\
        \hline
         64 & 38 & 16 & - & - & 10 \\
        \hline
        128 & 55 & 36 & 27 & - & 10 \\
        \hline
        256 & 76.3 & 59.4 & 42.4 & 25.5 & 52.4 \\
        \hline\hline
    \end{tabular}
\end{table}

In this section, we provide an overview of the IVAS codec in SBA mode.
It will be useful to discuss the results of the experiments.

The IVAS codec in SBA mode~\cite{weckbecker_ambisonics_2025} involves a combination of SPAR~\cite{mcgrath_immersive_2019} and DirAC~\cite{laitinen_reproducing_2011, hold_perceptually-motivated_2024}.
These parametric approaches rely on the extraction of spatial parameters with a low temporal resolution (from 5 to 20~ms) and the transmission of a limited number of audio channels (from 1 to 4).
The transport channels (TCs) are coded independently as mono signals with an EVS codec, called core-codec.
To make the coding as efficient as possible, the strategy consists in minimizing the information redundancy between the TCs, \ie, the correlation between them.
We choose to describe only SPAR as it is the algorithm mainly used to process the bandwidth below 4.4~kHz and as DirAC can be interpreted as a simplified version of SPAR.
For further details, \cite{weckbecker_ambisonics_2025} describes the SPAR and DirAC combination and \cite{noauthor_3gpp_2023} provides a comprehensive description of the IVAS codec.

First, a filter bank is applied to the signal to enable processing in the time-frequency (T-F) domain.
The $N$ input HOA components, denoted $\xv\in\mathbb{C}^N$ for all T-F bins, are divided into three channel subsets $\xv = \left[w, \xv_\text{r}^T, \xv_\text{p}^T \right]^T$ with $w$ the omnidirectional component, $\xv_\text{r}$ the $M\leq3$ directional components whose residual is transmitted to the decoder, and finally $\xv_\text{p}$ the $K$ non-transmitted components which will be reconstructed from the TCs and the spatial parameters.
The first TC is defined as:
\begin{equation}
    w' = \av^T . \xv_\text{FOA},
\end{equation}
where $\xv_\text{FOA}$ contains the first four elements of $\xv$ and $\av\in\mathbb{R}^{4}$ can be interpreted as a beamforming vector allowing to gather the energy in $w'$ and contributing to the decorrelation of the TCs.
The others TCs are defined as follows:
\begin{equation}
    \xv_\text{r}' = \xv_\text{r} - \pv_\text{r} . w',
\end{equation}
with $\pv_\text{r}\in\mathbb{R}^M$ a vector predicting the contribution of $w$ in $\xv_\text{r}$, estimated from the correlation.
The residual of the non-transmitted components is defined as:
\begin{equation}
    \xv_\text{p}' = \xv_\text{p} - \pv_\text{p} . w' - \Cm . \xv_\text{r}',
\end{equation}
where $\pv_\text{p}\in\mathbb{R}^K$ and $\Cm\in\mathbb{R}^{K\times M}$ predicting the contribution of $w'$ and $\xv_\text{r}'$ in $\xv_\text{p}$, respectively.
The energy of each component is transmitted to the decoder as a vector $\dv\in\mathbb{R}_+^K$.
Then, the TCs and spatial parameters are coded before the transmission to the decoder.
The number of TCs, $M+1$, varies depending on the total bitrate.
The bitrate allocation is described in Table~\ref{tab:ivas_bitrate_summary}.
While the TCs are coded independently by an EVS core-codec, spatial parameters ($\pv_\text{r}$, $\pv_\text{p}$, $\Cm$ and $\dv$) are frequency-downsampled and approximated by a non-uniform scalar quantizer.
In the following, for a variable $a$, we denote its coded/decoded version $\hat{a}$.

At the decoding stage, the reconstructed signal, denoted $\hat{\xv}=\left[\hat{w}, \hat{\xv_\text{r}}^T, \hat{\xv}_\text{p}^T \right]^T$, is defined as follows:
\begin{equation}
    \begin{bmatrix}
        \hat{w} \\
        \hat{\xv}_\text{r} \\
        \hat{\xv}_\text{p} \\
    \end{bmatrix}
         = \begin{bmatrix*}[l]
            g_\text{w} . \hat{w}' \\
            \hat{\pv}_\text{r} . \hat{w}' + \hat{\xv}_\text{r}' \\
            \hat{\pv}_\text{p} . \hat{w}' + \hat{\Cm} . \hat{\xv}_\text{r}' + \hat{\dv} \circ \begin{bmatrix} \mathcal{D}_1(\hat{w}') \\
																											\vdots \\
																											\mathcal{D}_K(\hat{w}') \\
																											\end{bmatrix}
            \end{bmatrix*},
\end{equation}
where $g_\text{w}$ is a gain correcting the effect of $\av$ during the encoding, and $\mathcal{D}_i(.)$ is a function generating a decorrelated version of the signal from its input for the $i$\textsuperscript{th} parametrized HOA component.
This function allows resynthesizing from $\hat{w}'$ the part of the signal which could not be predicted from the TCs.
Finally, $\hat{\xv}$ is converted back to time domain using an inverse filter bank.

\section{Experiment 1}\label{sec:exp1}

The aim of this experiment is to test the performance of the IVAS and EVSx16 codecs in terms of global quality on a wide variety of contents.
Particular attention is paid to including items derived from various and well-controlled spatialization methods.

\subsection{Methods}\label{ssec:exp1_methods}  

\begin{table}[b]
    \small
    \centering
    \caption{\label{tab:items}Summary of the list of items. Without further notice, items are original content.}
    \begin{tabular}{r | l l l}
        \hline\hline
        \# & Item  & Type & Spatialization \\
        \hline\hline
         1 & AMB        & Party~\cite{noauthor_clarity_2024}    & EM64 recording \\
         2 & APP        & Applause                              & EM32 recording \\
        \hline
         3 & FLK\_ANE   & Folk music                            & Ideal plane wave \\
         4 & FLK\_REV   & Folk music                            & ZM-1 recording \\
         5 & BND        & Jazz quartet                          & EM32 recording \\
         6 & ORC        & Orchestra                             & EM32 recording \\
         7 & POP        & Pop song                              & Ideal plane wave \\
        \hline
         8 & SPK1\_ANE  & 1 speaker~\cite{dubey_icassp_2022}    & Ideal plane wave \\
         9 & SPK3\_ANE  & 3 speakers~\cite{dubey_icassp_2022}   & Ideal plane wave \\
        10 & SPK1\_REV  & 1 speaker~\cite{dubey_icassp_2022}    & EM32 SRIR \\
        11 & SPK3\_REV  & 3 speakers~\cite{dubey_icassp_2022}   & EM32 SRIR \\
        12 & THE        & Drama                                 & EM32 recording \\
        13 & MTG        & Meeting                               & EM32 recording \\
        \hline\hline
    \end{tabular}
\end{table}

\begin{figure}[t]
    \begin{center}
        \includegraphics[width=\columnwidth]{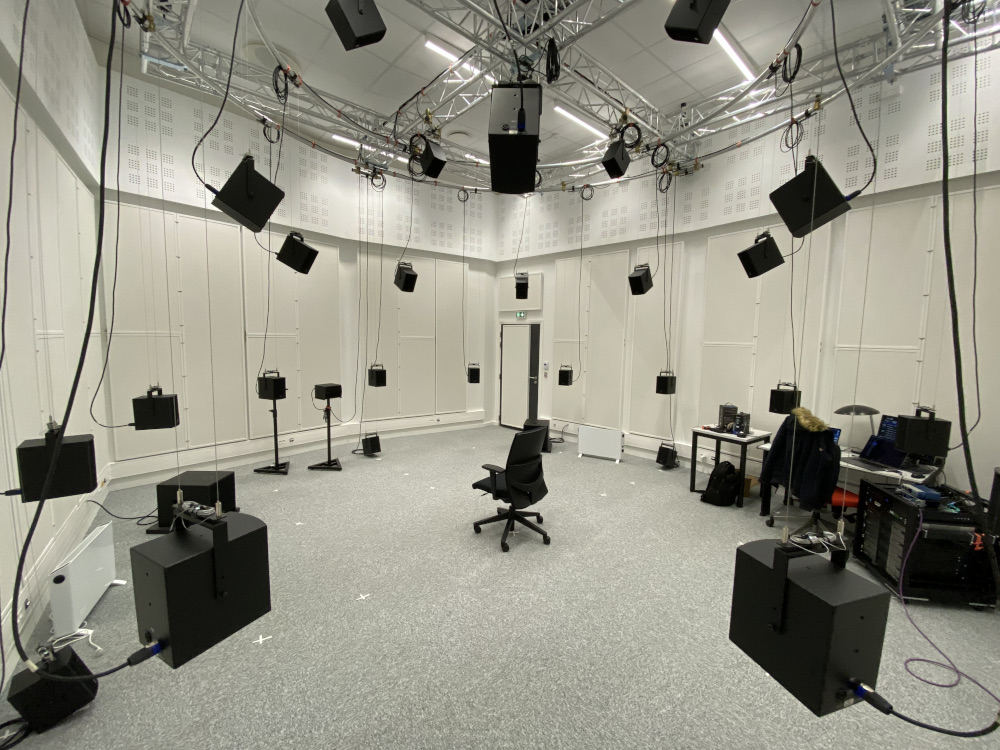}
        \caption{Listening room picture.}
        \label{fig:varese}
    \end{center}
\end{figure}

\begin{table}[b]
    \small
    \centering
    \caption{\label{tab:cut}Summary of the conditions under test.}
    \begin{tabular}{l | l r}
        \hline\hline
        Name  & Description & \makecell{Bitrate\\ (kbps)} \\
        \hline\hline
        Reference   & Uncompressed signal               & 12.3k \\
        \hline
        Anchor 3.5k & \makecell[l]{Low-pass filtered\\ reference at 3.5~kHz} & 12.3k \\
        \hline
        IVAS 32     & IVAS codec in SBA mode            & 32 \\
        \hline
        IVAS 64     & IVAS codec in SBA mode            & 64 \\
        \hline
        IVAS 128    & IVAS codec in SBA mode            & 128 \\
        \hline
        IVAS 256    & IVAS codec in SBA mode            & 256 \\
        \hline
        EVSx16 262    & \makecell[l]{EVS mono codec at 16.4~kbps\\ applied to each channel} & 262 \\
        \hline
        EVSx16 512    & \makecell[l]{EVS mono codec at 32~kbps\\ applied to each channel} & 512 \\
        \hline\hline
    \end{tabular}
\end{table}

A subjective listening test following the Multi-Stimuli with Hidden Reference and Anchors (MUSHRA, BS.1435) protocol~\cite{noauthor_itu-r_2015} is carried out using a loudspeaker dome.
In the following, we detail the choice of presented items and codecs, the participants as well as the listening setup.

\subsubsection{Items}  
The items are selected to be as varied as possible, including speech (one or three overlapping speakers), music and ambiances.
The spatialization techniques are also diverse: ideal plane wave encoding, convolution by SRIR of SMA, and native recording with SMA.

For speech, several degrees of realism and complexity are gradually tested for a codec.
One (\emph{resp.} three) speech sample from~\cite{dubey_icassp_2022} is spatialized using an ideal plane wave encoding (\emph{resp.} SPK1\_ANE and SPK3\_ANE), as well as convolved by SRIR (\emph{resp.} SPK1\_REV and SPK3\_REV) measured with an EM32 in an office (RT60 = 0.51~s).
Finally, we consider native recordings of auditory scenes captured with an EM32: a theatrical scene (THE) and a meeting (MTG) between three participants in a low-reverberation office.

For music, we consider two recordings captured with an EM32: a symphonic orchestra (ORC) and a small jazz band (BND).
For a simpler scenario and to study specifically the influence of spatially-diffuse reverberation, we used a Zylia ZM-1 to record a scene (FLK\_REV) with two instruments (violin and banjo) playing a simple tune in unison located at azimuth $\pm 45^\circ$ on the horizontal plane at a distance of 2~m.
By recording instruments separately, the omnidirectional component of each recording can be isolated and spatialized in the original direction using an ideal plane wave encoding (FLK\_ANE).
Finally, we consider a recording of pop music made of a female voice and three cellos spatialized with an ideal plane wave encoding (POP).

For the complex ambiances, we select a sample of indoor party background from~\cite{noauthor_clarity_2024} recorded with an EM64 (AMB).
We also use an applause recorded with an EM32 in a large music hall (APP).

Each item has a duration of approximately 10~s and a global level of -30~LUFS~\cite{noauthor_ebu_2023}, sampled at 48~kHz and limited to the 3\textsuperscript{rd}-order ambisonic.
Table~\ref{tab:items} summarizes the items content and associated spatialization method.

\begin{figure*}[t]
    \begin{center}
        \includegraphics[width=.9\textwidth]{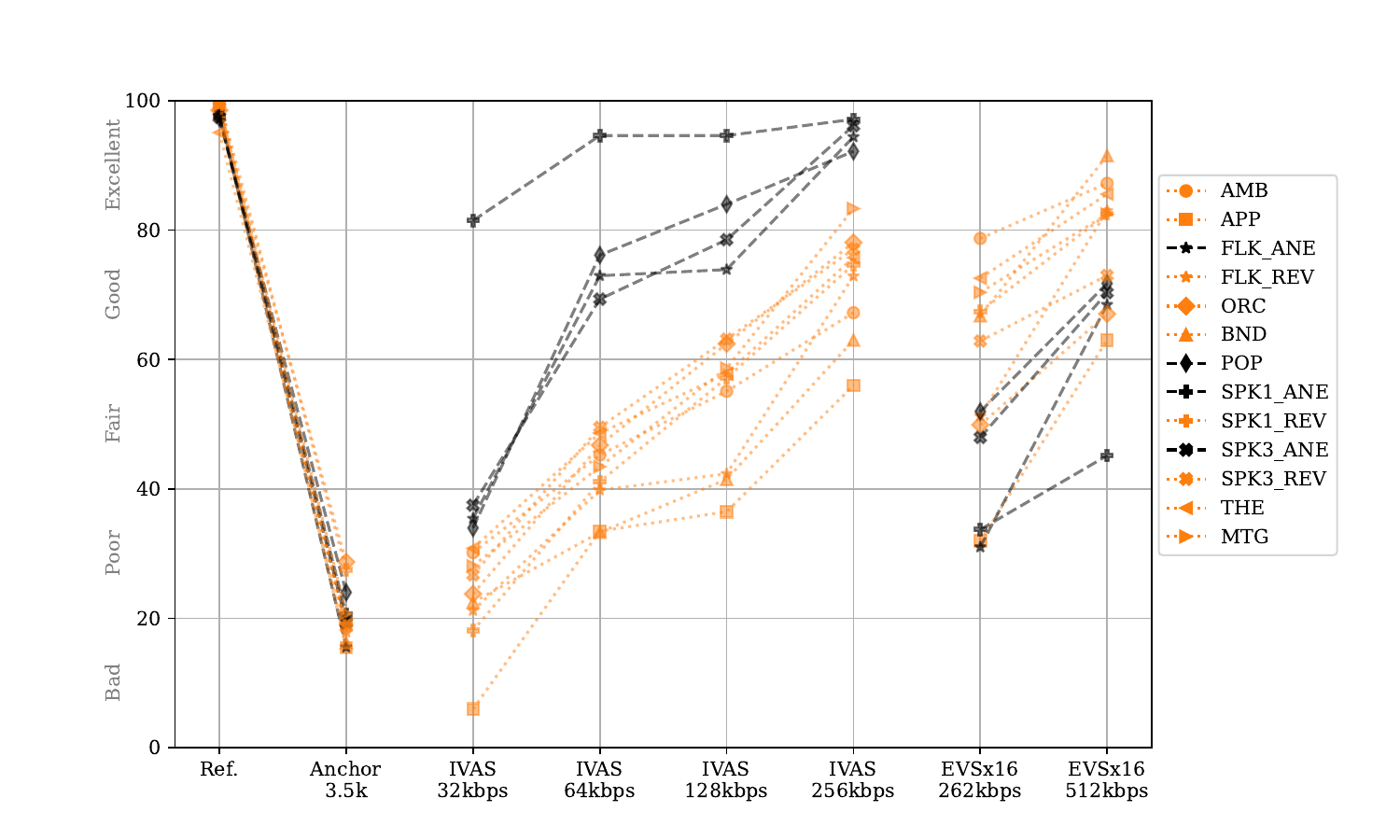}
        \caption{MUSHRA scores averaged across listeners for each listening condition and item. The items spatialized with an ideal plane wave encoding are set in black, the others (convolved by SRIR and recorded by an SMA) are set in orange.}
        \label{fig:mushra_mean_conditions_items}
    \end{center}
\end{figure*}

\begin{figure*}[t]
    \begin{center}
        \includegraphics[width=\textwidth]{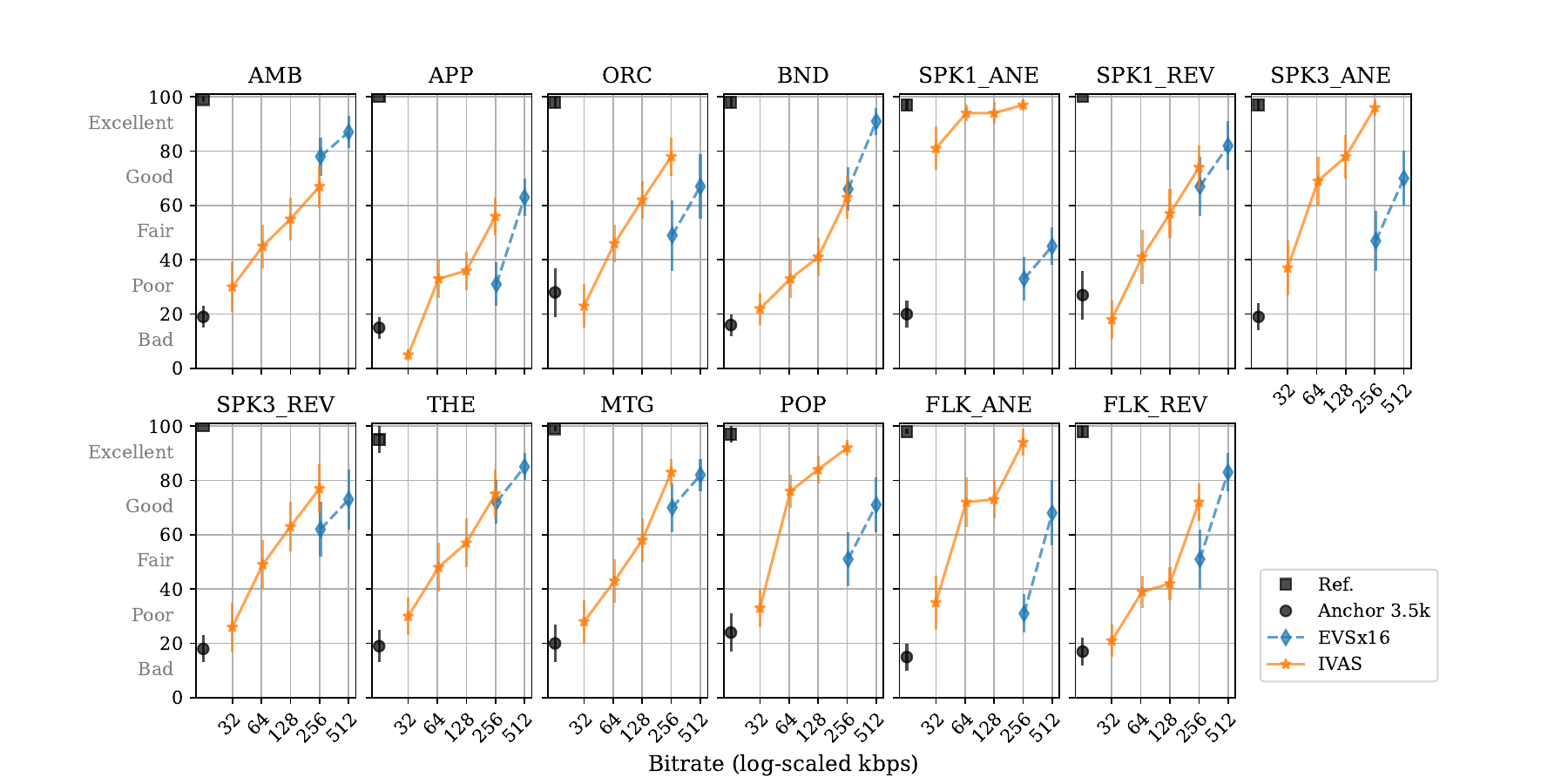}
        \caption{MUSHRA scores averaged across listeners for each listening condition and item. The vertical segments show the mean confidence interval at 95~\%. The hidden reference and the low anchor are shown on the left without indication of bitrate.}
        \label{fig:mushra_mean_ci95_conditions_items}
    \end{center}
\end{figure*}

\subsubsection{Codecs}  
For the conditions under test (CuT), we consider:
(i) the IVAS codec in SBA mode~\cite{noauthor_codec_2024, weckbecker_ambisonics_2025} at bitrates 32, 64, 128, and 256~kbps,
(ii) the EVS codec~\cite{dietz_overview_2015} applied independently to each HOA component at bitrates 16$\times$16.4~kbps and 16$\times$32~kbps, similarly as in~\cite{narbutt_streaming_2017} with Opus or in \cite{weckbecker_ambisonics_2025} with EVS at the 1\textsuperscript{st}-order ambisonic,
(iii) a low-pass filtered anchor at 3.5~kHz.
We focus on these four IVAS bitrates because the number of TCs increases incrementally from 1 to 4~\cite{noauthor_3gpp_2023, weckbecker_ambisonics_2025}.
See Table~\ref{tab:ivas_bitrate_summary} for details on the binary allocation for each TC and the spatial parameters.
The highest IVAS bitrates (384 and 512~kbps) are not tested in this study in order to keep the number of CuT reasonable.
We choose to use EVSx16 as the baseline rather than Opus CMF~2, since IVAS uses EVS as its core codec.
This limits bias in the interpretation of results by ensuring that observed differences are due to the spatial coding method, rather than artifacts from different core-codecs.
Additionally, we opted for a multi-mono approach for EVS, similar to Opus CMF~2 rather than CMF~3, for its simplicity and because \cite{rudzki_improvements_2023} showed that the quality improvement brought by the latter was not clear.
Nevertheless, it should be noted that Opus CMF~2 uses a stereo Opus core-codec, not a mono one.
Table~\ref{tab:cut} summarizes the CuT.

\subsubsection{Participants}  
The experiment was conducted with 19 subjects aged between 24 and 57 years (mean age: 40 years; 12 males and 7 females).
All were experts with prior experience in similar listening tests.

\subsubsection{Setup}  

Although headphones are the preferred device for listening to spatial audio in AR/VR contexts, we choose to conduct the listening test using a loudspeaker system.
This setup avoids issues related to (non)-individualized binaural listening~\cite{rudzki_auditory_2019} and artifacts associated with the choice of binaural rendering methods for ambisonic components~\cite{schorkhuber_binaural_2018, berebi_ambisonics_2025}.
Nevertheless, this approach requires selecting a spatial decoding method.
We opted for a widely used and readily available technique: the AllRADecoder plug-in from the IEM Suite~\cite{iem_iem_2021}, with maxrE weighting~\cite{zotter_all-round_2012}.

The listening test took place in a listening studio of 40~m$^2$ with a ceiling height of 4~m, see \figref{fig:varese}.
Both the background noise below the NR15 curve and the RT$_{60}$ of 0.29~s comply with the ITU-R~BS.1116~\cite{noauthor_itu-r_2015-1}.
The signals are played through a dome of 29 Amadeus PMX5 loudspeakers arranged with a radius of 2.7~m.
The loudspeakers are evenly distributed on horizontal rings at five different elevations: 4 at -22,\kern-\wd\commabox\degree \ 12 at 0,\kern-\wd\commabox\degree \ 8 at 25,\kern-\wd\commabox\degree \ 4 at 60,\kern-\wd\commabox\degree \ and 1 at 90.\kern-\wd\commabox\degree \
For each ring, one speaker is placed at azimuth 0.\kern-\wd\commabox\degree \
The system is calibrated with a Trinnov Optimizer processor compensating the delays and the frequency response.
The global gain is set to ensure that 65~dB(A) corresponds to -30~LUFS for spatialized pink noise in the frontal direction at the listening position.
Listeners can freely adjust the level by $\pm$~4dB and rotate on the chair with no translation allowed.

\subsection{Results}  

\begin{figure}[t]
    \begin{center}
        \includegraphics[width=\columnwidth]{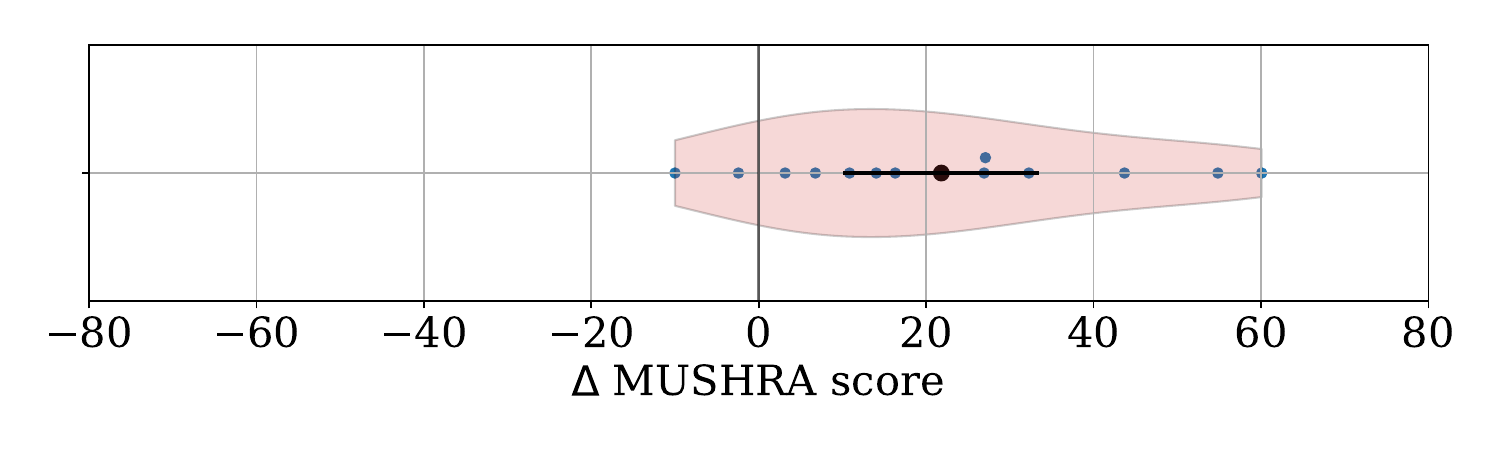}
        \caption{Difference of the listeners-averaged MUSHRA score between IVAS (at 256~kbps) and EVSx16 (at 262~kbps) for each item (small dots) and the overall mean (black dot) with its confidence interval at 95~\% (black segment). The violin plot illustrates the underlying distribution.}
        \label{fig:dmushra_ivas_vs_evs}
    \end{center}
\end{figure}

The results are reported in \figref{fig:mushra_mean_conditions_items} as MUSHRA scores averaged across listeners for each item and each CuT.
For readability, no indicator of data dispersion is shown in this figure.
In \figref{fig:mushra_mean_ci95_conditions_items}, we present detailed MUSHRA results for each item showing the mean and its 95~\% confidence interval.
In \figref{fig:dmushra_ivas_vs_evs}, we report the difference in MUSHRA scores between IVAS and EVS at the only comparable bitrate (256 and 262~kbps, respectively).
The hidden reference is clearly identified, and the 3.5~kHz anchor is consistently rated as poor quality for most items.

In \figref{fig:mushra_mean_conditions_items}, the robustness of IVAS performance can be observed for bitrates $\geq64$~kbps for the SPK1\_ANE signals and, to a lesser extent, POP, SPK3\_ANE, and FLK\_ANE.
In particular, speech alone spatially encoded as a plane wave (SPK1\_ANE) maintains excellent quality ($>80$), and this remains true even at the lowest bitrate (32~kbps).

For IVAS, two clusters of items with different behaviors can be identified.
The first, gathering signals spatialized with ideal plane wave encoding, maintains good quality ($>60$) for bitrates from 64 to 256~kbps.
The second, consisting of items with a spatially diffuse component, shows reduced performance compared to the first cluster at the same bitrate.
Conversely, EVSx16 seems to perform better for signals with spatially diffuse reverberation than for those generated with ideal plane wave spatial encoding.

In \figref{fig:mushra_mean_ci95_conditions_items}, it is noted that for IVAS, increasing the bitrate leads to a quality gain.
However, a stagnation in quality is observed between 64 and 128~kbps for certain items: APP, SPK1\_ANE, FLK\_ANE, and FLK\_REV.

Our experimental protocol allows to compare IVAS to EVSx16 at a similar bitrate around 256~kbps (256 for IVAS, 262 for EVSx16).
In \figref{fig:dmushra_ivas_vs_evs}, it can be seen that the mean for IVAS is consistently higher than that of EVSx16 for all items except AMB and, to a lesser extent, BND, where the mean difference is not statistically significant.
For AMB, the difference is 11 points in favor of EVSx16.

\section{Experiment 2}\label{sec:exp2}

In this second experiment, we specifically investigate the effect of spatially diffuse reverberation on the quality of signals degraded by the considered codecs.
This particularly relates to hypotheses \hyp{1} to \hyp{4} presented in the introduction.

\subsection{Methods}  

We essentially reuse methodological elements from the first experiment by replicating similar MUSHRA test sessions in order to increase the statistical power of the results.
Without further notice, we consider the method and setup described in Section~\ref{sec:exp1}.

\subsubsection{Items}  

To test \hyp{1} and \hyp{3} with all other factors held constant, we synthesize auditory scenes with mono signals that are spatialized either with ideal plane wave spatial encoding or by convolution with SRIRs in the same directions.
Details are explained in Section~\ref{sec:exp1} for the pairs of signals SPK1\_ANE and SPK1\_REV, SPK3\_ANE and SPK3\_REV, and FLK\_ANE and FLK\_REV.
In total, we generate five new test sessions following the same model as described for Experiment~1, each including 13~items.
However, the results for items other than SPK1, SPK3, and FLK in their ANE and REV versions are not further analyzed.
It leads to six test sessions in total (including the one from the first experiment) for each subject, of approximately 1~hour each.
The order of the five additional test sessions is randomized for each subject.

\subsubsection{Participants}  

For this experiment, we consider a subset of $L=6$ participants who took part to the first one (4 males and 2 females, average of 38~years).

\subsubsection{Criterion and statistical analysis}

The analysis of the results is limited to the signal classes SPK1, SPK3, and FLK, for which there is a pair of signals: either consisting of a limited number of plane waves (with the \_ANE suffix) or spatialized with a measured SRIR (with the \_REV suffix).
We define the diff-MUSHRA score as the difference of score between anechoic and the corresponding spatially-reverberated MUSHRA scores averaged across listeners, denoted $\Delta \bar{s}_i$:
\begin{equation}
    \Delta \bar{s}_i = \frac{1}{L} \sum\limits_{\ell=1}^L s^{(a)}_{i, \ell} - s^{(r)}_{i, \ell},
\end{equation}
with $i$ the item index, $\ell$ the listener index, and where $s^{(a)}$ and $s^{(r)}$ are the MUSHRA scores of the anechoic version of the item and its spatially-reverberated counterpart, respectively.

We need to test the statistical significance of the difference between the mean scores between two listening conditions (codecs).
As our data are paired, we use a paired sample $t$-test with a threshold of 0.05.
As we compare multiple pairs of listening conditions, we need to control the false discovery rate at 5~\%.
To do so, we apply the Benjamini-Hochberg $p$-value adjustment.
Prior to further analysis, a Shapiro-Wilk test is used to verify that the data meet the normality assumption.

\subsection{Results}  

\begin{figure}[t]
    \begin{center}
        \includegraphics[width=\columnwidth]{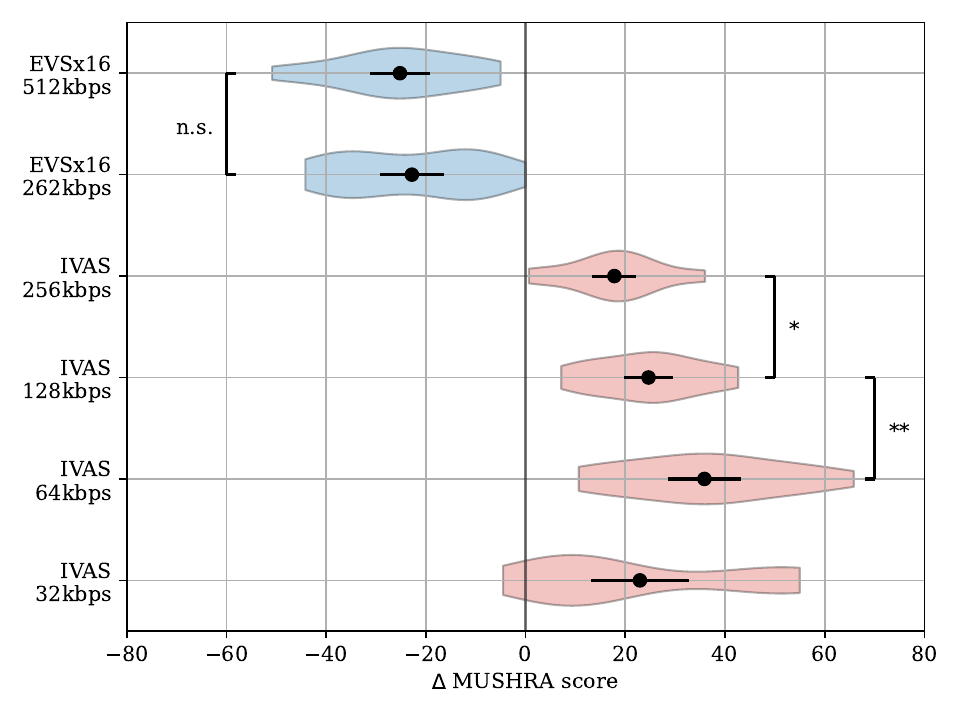}
        \caption{For each codec, the listener-averaged MUSHRA scores differences between anechoic and the corresponding spatially-reverberated 18 items (3 classes, 6 sessions). The black dot denotes the mean, the black segment denotes the 95~\% mean confidence interval. The underlying distribution is illustrated thanks to a violin plot. \emph{n.s.} denotes a non-significant difference between two conditions, \emph{\textsuperscript{*}} denotes a significance $p<0.05$ and \emph{\textsuperscript{**}} denotes a significance $p<0.01$.}
        \label{fig:dmushra_anech_reverb}
    \end{center}
\end{figure}

The results are summarized in \figref{fig:dmushra_anech_reverb} for each codec, shown as the mean and 95~\% confidence interval of the mean, as well as a violin plot to illustrate the empirical distribution of MUSHRA score differences.

First, it is observed that EVSx16 shows a bias in favor of reverberant signals ($\Delta \bar{s}<0$) of more than 20 MUSHRA points, giving credence to \hyp{3}.
This bias is not depending on bitrate ($p=0.747$) on the tested range, supporting \hyp{4}.

Next, an opposite effect is observed for IVAS, with a bias in favor of anechoic signals of 20 points at a bitrate of 256~kbps, and up to 35 points at 64~kbps, which supports \hyp{1}.
For the three highest tested bitrates (64, 128, and 256~kbps), a decreasing effect of the bias as a function of bitrate is observed ($p=0.013$ for IVAS 256 vs. 128 and $p=0.002$ for IVAS 128 vs. 64).
Finally, for IVAS at 32~kbps, a bias in favor of anechoic signals is still observed, but it does not follow the decreasing trend seen at higher bitrates.
As shown in \figref{fig:mushra_mean_conditions_items} for IVAS at 32~kbps, this difference results from the collapse of MUSHRA scores for anechoic signals, rather than from an improvement in the quality of reverberant signals.
This result supports \hyp{2} with a more complex behavior observed at the lowest bitrates ($\leq$ 32~kbps).

\section{Discussion}\label{sec:discussion}

In line with previous work~\cite{noauthor_3gpp_2024, weckbecker_ambisonics_2025}, our results show that IVAS provides a quality gain compared to a naive ambisonic coding approach, EVSx16, which is comparable to Opus CMF~2, at similar bitrates.
Indeed, EVSx16 cannot explicitly exploit and preserve inter-channel correlation, reducing the spatial reconstruction quality.
Our results identified that artifacts are particularly audible on signals whose HOA components are originally highly correlated, see \figref{fig:dmushra_anech_reverb}.
Thus, instead of perceiving a clearly localized sound source, it is perceived as very diffuse or spread across multiple locations simultaneously.
More detailed informal listening revealed that artifacts are concentrated on transient sounds or speech fricatives.
This is typical of a core-codec like EVS, which does not faithfully reconstruct the phase of the signal in some cases at lower bitrates~\cite{dietz_overview_2015}.
Conversely, this lack of inter-channel correlation handling is less detrimental for highly reverberant signals or those capturing a complex sound scene, even though IVAS is almost always better at equal bitrate, see \figref{fig:dmushra_ivas_vs_evs}.

In addition to spatialization artifacts, the multiplication of independently coded channels forces the bitrate to be distributed among channels that are sometimes highly redundant, to the point of using EVS for each TC at a very low bitrate where coding artifacts affecting timbre become audible.
Specifically, the per-channel bitrate of EVSx16 at 262~kbps (16.4~kbps/channel) should be compared to the bitrates for the four TCs transmitted by IVAS at 256~kbps, see Table~\ref{tab:ivas_bitrate_summary}.
Thus, by grouping information into a limited number of TCs, IVAS can use higher bitrates per TC and thereby reduce timbre artifacts.

A more in-depth analysis highlighted a significant disparity in IVAS performance depending on the items.
In particular, we identified the robustness of IVAS performance on signals composed of a limited number of ideal plane waves, even at low bitrates.
We attribute this to the fact that IVAS relies on a combination of algorithms, SPAR and DirAC, which strongly exploit inter-channel correlation to compress information, as described in Section~\ref{sec:ivas}.
Their performance drops on reverberant signals.
Indeed, it has been shown that reverberation leads to decorrelation of HOA components~\cite{hellerud_spatial_2009}, making compression more difficult.
Moreover, both experiments showed that the degradation of IVAS performance on spatially-diffuse reverberation depends on the bitrate.
This performance drop can be explained by two main, mutually non-exclusive, hypotheses.
On the one hand, IVAS relies on decorrelation filters to reproduce the spatially-diffuse part of the signal not transmitted by the TCs.
However, as the bitrate decreases, IVAS reduces the number of TCs, and thus relies more on decorrelation filters for reconstructing this diffuse part.
The mismatch between these filters and the reverberation to be reproduced may explain the perceived quality loss.
On the other hand, the reduction of IVAS bitrate impacts the bitrate of the TCs, resulting in more core-codec artifacts.
To individually test these hypotheses, full control over the algorithms would be required, which is beyond the scope of this study.

Nevertheless, an indirect analysis of our results supports the role of decorrelation filters in the degradation of IVAS performance.
Indeed, Experiment~1 highlighted a stagnation in quality gain between 64 and 128~kbps for the items SPK1\_ANE and FLK\_ANE, see \figref{fig:mushra_mean_ci95_conditions_items}.
These consist of one and two plane waves, respectively.
It is therefore expected that adding a third TC does not improve quality\footnote{Conversely, SPK3\_ANE and POP, composed of three plane waves, benefit from the third TC to increase quality.}.
However, this result also shows that increasing the bitrate per TC does not provide a quality gain in this range and for these signals.
This tends to refute the second hypothesis.
On the other hand, SPK1\_REV does benefit from the increase from 64 to 128~kbps to improve perceived quality, which can thus be attributed to the addition of a third TC.
For FLK\_REV, whose reverberation is less pronounced, it is observed that the addition of a third TC only slightly improves the coding of spatially diffuse reverberation.
This result suggests that, in the presence of diffuse reverberation, the IVAS decorrelation filters poorly substitute for the addition of a TC.
A more in-depth study of these cases is needed to support or invalidate these hypotheses.

Ultimately, we have highlighted a bias of the SPAR and DirAC algorithms, combined in IVAS in SBA mode, in favor of content with strong inter-channel correlation, \ie, signals without a spatially diffuse component, \eg, composed of a limited set of plane waves.
This result points to a research direction for the next generation of codecs, where further effort is needed to improve the quality of spatially diffuse signals at low bitrates, \ie, including natural reverberation or complex ambience, without losing quality on non-diffuse signals.


\bibliographystyle{ieeetr}
\bibliography{bible}

\end{document}